\documentstyle[aps]{revtex}
\draft
\input epsf
\begin{document}
\twocolumn[\hsize\textwidth\columnwidth\hsize\csname@twocolumnfalse%
\endcsname
\title{Quantum Force in Superconductor}
\author{A. \ V. \ Nikulov}

\address{Institute of Microelectronics Technology and High Purity
Materials, Russian Academy of Sciences, 142432 Chernogolovka, Moscow
District, RUSSIA}

\maketitle
\begin{abstract}
{In order to conciliate a contradiction of the Little-Parks experiment with
the Ohm's law and other fundamental laws the thermal fluctuation is
considered as dynamic phenomenon and a new force called "quantum force" is
introduced.  A persistent current can exist at zero voltage and non-zero
resistance because of the quantum force induced by the thermal fluctuation.
Not only the persistent current but also a persistent voltage (a direct
voltage in the equilibrium state) can exist in an inhomogeneous
superconducting ring.  Directions of the persistent current and the
persistent voltage coincide in a ring segment with lower critical
temperature and are opposite in other ring segment with higher $T_{c}$.
Consideration of a superconducting ring interrupted by Josephson junction
shows a connection of the quantum force with a real  mechanical force.}
\end{abstract}

 ]

\narrowtext

Superconductivity is a macroscopic quantum phenomenon: some macroscopic
effects observed in superconductors can not be described by classical
mechanics.  One of them is the Little-Parks  (LP) experiment
\cite{little62} repeated in numerous works (see for example \cite{repeat}).
It is considered  \cite{tink75,repeat} that the LP experiment was explained
as long ago, as 1963 \cite{tinkham}. But this explanation is not perfect.
More perfect consideration shows a contradiction of the LP experiment in a
loop \cite{repeat} with some habitual knowledge. This contradiction is
explained in the present work.

The resistance oscillations observed at the LP experiment is interpreted as
a consequence of oscillations of the superconducting transition temperature
$T_{c}$ \cite{little62,tink75,repeat}.  It is assumed \cite{tink75} that
$R_{l}(T) = R_{l}(T-T_{c})$ in the resistive transition region where the
resistance  $R_{l} = \int_{l}dl \rho/s$  \cite{repeat} changes from $R_{l}
= R_{ln}$ to $R_{l} = 0$.  Here $R_{ln}$ is the resistance along the loop
in the normal state; $s$ is the area of the cross-sectional of the wire
defining the loop.

The  $T_{c}$ oscillations is explained by the fluxoid quantization
\cite{tinkham,tink75,repeat}. Because of the quantization the velocity
circulation $$\int_{l}dl v_{s} = \frac{\pi \hbar}{m} (n
-\frac{\Phi}{\Phi_{0}}) \eqno{(1)} $$ of superconducting pairs can not be
equal zero when the magnetic flux  $ \Phi$ contained within a loop is not
divisible by the flux quantum $\Phi_{0} = \pi \hbar c/e$. Therefore the
energy of superconducting state increases and as consequence the  $T_{c}$
decreases  when  $ \Phi \neq n\Phi_{0}$, $\Delta  T_{c} \propto -v_{s}^{2}
\propto  -(n -\Phi/\Phi_{0})^{2}$ \cite{tink75}. The magnetic flux $LI_{s}$
induced by the screening current  $I_{s} = sj_{s} =s 2en_{s}v_{s}$ is small
$LI_{s} \ll \Phi_{0}$  at  $T \simeq T_{c}$  (when  the density of
superconducting pairs $n_{s}$ is close to  zero) and therefore  $\Phi = BS
+ LI_{s}\simeq BS$ \cite{tink75}. Here $B$ is the magnetic induction
induced by an external magnet; $S$ is the area of the loop.

It is important that the theoretical dependence  $\Delta  T_{c} \propto
-(n -\Phi/\Phi_{0})^{2}$, where $v_{s}^{2} \propto (n -\Phi/\Phi_{0})^{2}$
has minimum possible value \cite{tink75} describes enough well the
experimental data (see for example Fig.4 in  \cite{repeat}).  Consequently,
superconducting states with minimum $(n -\Phi/\Phi_{0})^{2}$ value give the
main contribution. This means that not only the average
$\overline{v_{s}^{2}} = t_{long}^{-1} \int_{t_{long}}dt v_{s}^{2}$ but also
$\overline{v_{s}} = t_{long}^{-1} \int_{t_{long}}dt v_{s}$ is not equal
zero at $ \Phi \neq n\Phi_{0}$ and $ \Phi \neq (n+0.5) \Phi_{0}$.
$\overline{v_{s}} = 0$ at  $ \Phi = (n+0.5) \Phi_{0}$ because the permitted
states with opposite direction of the velocity have the same $v_{s}^{2}$
value. $\overline{v_{s}} \approx (\pi \hbar/ml) (n -\Phi/\Phi_{0})$ in a
homogeneous loop when $\Phi$ is not close to  $(n+0.5) \Phi_{0}$.

Thus, according to the LP experiment and in spite of the Ohm's law $R_{l}
I_{sc} = \int_{l}dl E = -(1/c)d\Phi/dt$ a direct screening current
$\overline{I_{sc}} \approx s2e\overline{n_{s}}(\pi \hbar/ml) (n
-\Phi/\Phi_{0})$ flows along the loop at a constant magnetic flux, $ \Phi
\neq n\Phi_{0}$ and $ \Phi \neq (n+0.5) \Phi_{0}$, and $R_{l} \neq 0$. The
later is evident from the experiment  \cite{repeat}. The measured
resistance  \cite{repeat} $R_{m} \approx R_{l}/4$ in a homogeneous loop.
The LP experiment contradicts not only to the Ohm's law  but also some more
fundamental laws because a dissipation (friction) force $F_{dis}$ should
act at $I_{sc} = sj_{sc}  \neq 0$ and  $R_{l} \neq 0$, and an energy
dissipation with power  $R_{l}I_{sc}^{2}$ should take place.

This contradiction has a explanation having a single meaning. It is obvious
that in a stationary state the screening current is equal superconducting
current $$j_{s}  = \frac{2e\pi \hbar}{lm<n_{s}^{-1}>}(n -
\frac{\Phi}{\Phi_{0}}) \eqno{(2)}$$ It can be non-zero when the whole of
loop in superconducting state, i.e.  $<n_{s}^{-1}>^{-1} \neq 0$, and $R_{l}
= 0$.  $<n_{s}^{-1}> = l^{-1}\int_{l}dl n_{s}^{-1}$ is used because the
$j_{s}$ value should be constant along the loop in the stationary state. (
$n_{s}$ ought be considered as an effective density in order to take into
account the Josephson current through segments with  $n_{s} = 0$).
Therefore the LP oscillations are observed only in the region of the
resistive transition where loop segments are switched by the thermal
fluctuation between superconducting state (when  $<n_{s}^{-1}>^{-1} \neq
0$, $j_{s}  \neq 0$ but $R_{l} = 0$) and normal state (when
$<n_{s}^{-1}>^{-1} = 0$, $R_{l} \neq 0$  but $j_{s} = 0$). These
oscillations can not be observed below the resistive transition where
$j_{s}  \neq 0$ but $R_{l} = 0$ all time and above this transition where
$R_{l} = R_{ln}$  but $j_{s} = 0$ all time.

Thus, the LP experiment is evidence of a motion induced by fluctuation in
the thermodynamic equilibrium state at non-zero dissipation. Such phenomena
are called Brownian motion \cite{Isihara}. There is an important difference
from the classical Brownian motion. According to the classical mechanics
the average velocity of any Brownian motion should be equal zero whereas
the LP experiment is evidence of the persistent current (i.e. a direct
current  in the equilibrium state) $j_{p.c.} = \overline{j_{sc}} \neq 0$ at
$ \Phi \neq n\Phi_{0}$ and $ \Phi \neq (n+0.5) \Phi_{0}$.

The persistent current $$j_{p.c.} = q\sum_{p}vf_{0}(\frac{E(p)}{k_{B}T}) =
\frac{q}{m}\sum_{p}(p - \frac{q}{c}A) f_{0}(\frac{E(p)}{k_{B}T})
\eqno{(3)}$$ is a quantum phenomenon. It can exist in states with discrete
spectrum $\int_{l}dl p = n2\pi \hbar$, at  the energy difference between
adjacent permitted states $E(n+1) - E(n) \geq  k_{B}T$, when the summation
$\sum_{p}$ can not be replaced by integration.  Here $p = mv + (q/c)A$ is
the generalized momentum of a particle with a charge $q$;  $A$ is the
vector potential. At continuous spectrum (at $E(n+1) - E(n) \ll  k_{B}T$)
$j_{p.c.} =  q\sum_{p}vf_{0} = q\int dv vf_{0} = 0$  because the
distribution function in the equilibrium state $f_{0}$ depends on $v$ only
through  $E(p)/k_{B}T$ and the kinetic energy is proportional to $v^{2}$ in
a consequence of the space symmetry.  Therefore, according to the classical
mechanics any direct (non-chaotic) current can be only in a nonequilibrium
state and it is postulated that the average value of the fluctuation force
introduced by Langevin for description of the classical Brownian motion is
equal zero $\overline{F_{Lan}} = 0$.

In a superconducting loop the difference between adjacent permitted states
of the kinetic energy $${\it E_{p}} = s\int_{l}dl
n_{s}\frac{2mv_{s}^{2}}{2}  = \frac{s\pi^{2} \hbar^{2}}{lm<n_{s}^{-1}>}(n -
\frac{\Phi}{\Phi_{0}})^{2} \eqno{(4)}$$ is proportional to
$<n_{s}^{-1}>^{-1}$ and of the energy of the magnetic flux induced by the
superconducting current ${\it E_{L}} = LI_{s}^{2}/2c^{2} =
(Ls^{2}e^{2}2\pi^{2}\hbar^{2}/c^{2}l^{2} m^{2}<n_{s}^{-1}>^{2}) (n -
\Phi/\Phi_{0})^{2} = (Ls/l\lambda _{0}^{2})n'_{s}E_{p}$ is proportional to
$<n_{s}^{-1}>^{-2}$.  Here $\lambda _{0} = (c^{2}2m/4e^{2}n_{s}(0))^{1/2}$
is the London penetration depth at $T = 0$; $n'_{s} =
(n_{s}(0)<n_{s}^{-1}>)^{-1}$; $n_{s}(0)$ is the density of superconducting
pairs at $T= 0$.  At weak screening, when the LP oscillations are observed,
$ (Ls/l\lambda _{0}^{2})n'_{s} < 1$ and consequently  ${\it E_{L}} < {\it
E_{p}}$.

Superconducting pairs, as condensed bosons, have the same value of the
momentum circulation $\int_{l}dl p = n2\pi \hbar$. Therefore the $E(n+1) -
E(n)$ value for superconducting pairs in a loop $l$ at $<n_{s}^{-1}>^{-1}
\approx <n_{s}> \neq 0$ is much more than the one for electron $E_{p}(n+1)
- E_{p}(n) = (2\pi^{2} \hbar^{2}/l^{2}m)[(n+1)^{2} - n^{2}] \approx
2\pi^{2} \hbar^{2}/l^{2}m$ because the average  number of superconducting
pairs $sl<n_{s}>$ is very big in a real case. For a real length $l \simeq 4
\mu m$ of the wire defining the loop \cite{repeat} $2\pi^{2}
\hbar^{2}/l^{2}m \simeq k_{B} \ 1 K$. Therefore  the persistent current in
normal metal mesoscopic systems \cite{Kulik} is observed only at very low
temperature \cite{IBM1991}.  In  superconductor the screening persistent
current $j_{p.c.} = q\sum_{p}vf_{qu} = j_{s}$ is observed even in
macroscopic samples (for example at the Meissner effect) because
$E_{p}(n+1) - E_{p}(n) \approx sl<n_{s}>(\pi^{2} \hbar^{2}/l^{2}m) \gg
k_{B}T$ even near $T_{c}$.

Consequently, in the region of the resistive transition the fluctuations
switch the loop between qualitatively different states: the superconducting
state $<n_{s}^{-1}>^{-1} \neq 0$ with strongly discrete spectrum
$|E_{p}(n+1) - E_{p}(n)| \gg k_{B}T$, in which the circulation of the
phase gradient  $\nabla \varphi = p/\hbar $ of the wave function of
superconducting pairs has a definite value $\int_{l}dl \nabla \varphi =
2\pi n$, and the state with continuous $p$ spectrum, in which
$<n_{s}^{-1}>^{-1} = 0$, $R_{l} \neq 0$, the energy $E_{p}(n) = 0$ for any
$n$ value and therefore  $\int_{l}dl \nabla \varphi$ is "bad" (vague)
number. The later means that the "random phase" assumption is valid at
$<n_{s}^{-1}>^{-1} = 0$ and therefore the average velocity should be equal
zero in the equilibrium state \cite{Kohn57}.

Thus, the average value of the momentum circulation of superconducting
pairs changes between   $\int_{l}dl p = \int_{l}dl (2mv_{s} + (2e/c)A) =
(2e/c)\Phi$ and $\int_{l}dl p = n2\pi \hbar$ at the switching between
$<n_{s}^{-1}>^{-1} = 0$ and $<n_{s}^{-1}>^{-1} \neq 0$. At $ (Ls/l\lambda
_{0}^{2})n'_{s} \ll 1$, when the $A$ change is small, the momentum change
on the unit volume $\Delta P \simeq (m/e) j_{s}$. These momentum changes
induced by fluctuations explain the contradiction of the LP experiment with
habitual laws. The persistent current $j_{p.c.} = \overline{j_{sc}} \neq 0$
can exist at non-zero dissipation $\overline{F_{dis}} \neq 0$ because  the
momentum circulation should return to the quantum value $n2\pi \hbar$ at
switching to the state with $<n_{s}^{-1}>^{-1} \neq 0$. The momentum
circulation does not change systematically during a long time $t_{long}$ at
$\int_{t_{long}}dt F_{dis} = t_{long}\overline{F_{dis}} \neq 0$ because at
reiterated switching $\int_{l}dl \overline{F_{dis}} + \int_{l}dl
\overline{\Delta P}\omega  = 0$.  $\overline{\Delta P} =
N_{sw}^{-1}\sum_{k} \Delta P(k)$; $\Delta P(k)$ is the momentum change at
$k$ switching in the state with $<n_{s}^{-1}>^{-1} \neq 0$;  $\omega  =
N_{sw}/t_{long}$; $N_{sw}$ is the number of switching for $t_{long}$.

At the closing of the superconducting state in the loop, as well as at the
Meissner effect, superconducting pairs are accelerated against the force of
the electric field $\int_{l}dl E = - (1/c) d\Phi/dt$. In order to eliminate
the contradiction with the Newton's law a force $F_{q}$ may be introduced,
$\overline{F_{q}} =  \overline{\Delta P}\omega  $. Because the $\Delta P$
is induced by quantization it is natural to call  $F_{q}$ as quantum force.
The necessity to introduce the  $F_{q}$ is conditioned by the well known
difference between superconductor and a classical conductor with infinite
conductivity. It is important that the quantum force can not be localized
in any segment of the loop in principle because of the uncertainty relation
$\Delta p \Delta l > \hbar $. The $v_{s}$ becomes non-zero when the
momentum takes a certain value, $\Delta p \ll p_{n+1} - p_{n} = 2\pi
\hbar/l $, i.e. when superconducting pairs  can not be localized in any
segment of the loop. $\overline{F_{q}}$ should be uniform along the loop
because $\Delta P \propto j_{s}$.

The quantum force $\overline{F_{q}}$ takes the place of the Faraday's
voltage $- (1/c) d\Phi/dt$ which maintains the screening current in a
conventional loop with $R_{l} \neq 0$. Therefore the $j_{p.c.} =
\overline{j_{sc}} \neq 0$ is observed at  $\overline{R_{l}} \neq 0$ and
$\overline{d\Phi/dt}$ in the LP experiment. The  periodic variation of the
resistance with magnetic field $R_{l}(\Phi/\Phi_{0})$ is observed in the LP
experiment  \cite{repeat} because the probability of superconducting state
$P(<n_{s}^{-1}>^{-1} \neq 0) \propto \exp -( {\it E_{p}} + {\it
E_{L}})/k_{B}T$  decreases at $\Phi \neq n\Phi_{0}$. The approximation
\cite{tink75}, in which only state with minimum $|n - \Phi/\Phi_{0}| $ is
taken into account, describes enough well the experimental data
\cite{repeat} because in the superconducting state  $|E_{p}(n+1) -
E_{p}(n)| \gg k_{B}T $ even in the fluctuation region near $T_{c}$.

Thus, the LP experiment is evidence of a direct (non-chaotic)
one-dimensional Brownian motion. The Brownian particle in this case is the
superconducting condensate. Its kinetic energy changes randomly in time:
the  ${\it E_{p}}$ (and also  ${\it E_{L}}$) is increased by the quantum
force and dissipates after the switching a loop segment in the normal
state. The quantum force induced by the fluctuations is the Langevin force
$F_{Lan}$. Contrary to the classical Brownian motion $\overline{F_{Lan}} =
\overline{F_{q}} = \overline{\Delta P} \omega   \neq 0$ at $j_{p.c.} \neq
0$ and $\overline{R_{l}} \neq 0$.

Because the LP experiment is explained by the fluctuation switching between
$j_{sc} = q\sum_{p}vf_{cl} = 0$, where the distribution function $f_{cl}$
is in the equilibrium $f_{0} = f_{cl}$ above $T_{c}$, and  $j_{sc} =
q\sum_{p}vf_{qu}\neq 0$ where $f_{qu}$ is in the equilibrium  $f_{0} =
f_{qu}$ below $T_{c}$, it is useful to consider  the motion along the loop
both superconducting pairs and electrons at the transition between $f_{cl}$
and $f_{qu}$. The reduction of $j_{sc}$ at  $R_{l} \neq 0$ can be described
by  the classical Boltzmann transport equation \cite{MetTheor} because the
"random phase" assumption is valid at  $<n_{s}^{-1}>^{-1} = 0$.  But the
$j_{sc}$ appearance contradicts classical mechanics. For a phenomenological
description  of the transition  $f_{cl} \rightarrow f_{qu}$,  a new term
$\aleph $ may be added to the Boltzmann equation $$\frac{df}{dt} =
\frac{\partial f}{\partial t} + v \frac{\partial f}{\partial l} + qE_{V}
\frac{\partial f}{\partial p} = \aleph  - \frac{f_{1}} {\tau} \eqno{(5)} $$
 $\aleph =  df/dt + f_{1}/\tau$ during a time  $\Delta t_{qu}$ of the
 transition  $f_{cl} \rightarrow f_{qu}$ and  $\aleph =  0$ during any
 other time, $\int_{\Delta t_{qu}} dt \aleph =  f_{qu} - f_{cl} +
 \int_{\Delta t_{qu}} dt f_{1}/\tau $.  $p$ is the generalized momentum.
 Therefore $E_{V}  = - \nabla V$ is the potential part of the electric
 field $E = -\nabla V - (1/c) \partial A/\partial t$: $qE_{V} = \partial
 p/\partial t = m \partial v/\partial t + (q/c)\partial A/\partial t = qE +
 (q/c)\partial A/\partial t = -q\nabla V$. The distribution function  $f =
 n_{s} + f_{e}$ describes both  superconducting pairs and electrons. $q =
 e$ for electron and  $q = 2e$ for superconducting pair. $f_{1} = f -
 f_{0}$ is the deviation of the distribution function $f$ from the one
 $f_{0}$ in the equilibrium state. It is assumed that the equilibrium
 distribution $f_{0} = f_{cl}$ at $<n_{s}^{-1}>^{-1} = 0$ and $f_{0} =
 f_{qu}$ at $<n_{s}^{-1}>^{-1} \neq 0$. The difference between  $f_{1} = f
 - f_{cl}$ and $f_{1} = f - f_{qu}$ is not important in our consideration
 because the mean time between collisions $\tau$ is infinite for
 superconducting pairs and the equilibrium distributions for electrons
 $f_{e}$ are approximately the same at $<n_{s}^{-1}>^{-1} = 0$ and
 $<n_{s}^{-1}>^{-1} \neq 0$.

The balance on the average forces $$\frac{\partial P}{\partial t} - F_{p} -
F_{e} = F_{q} - F_{dis} \eqno{(6)} $$ is obtained by multiplication of the
transport equation (5) by the momentum and summing over the $p$ states.
Here $P = \sum_{p}pf = p_{s}n_{s} + \sum_{p}pf_{e} = P_{s} + P_{e}$; $
P_{s}= p_{s}n_{s}$ is the momentum per unit volume of superconducting
pairs;  $P_{e} = \sum_{p}pf_{e}$  is the momentum per unit volume of normal
electrons; $F_{p} = - \partial (\sum_{p}pvf)/\partial l = - \partial
(n_{q}<pv>)/\partial l$ is the force of the pressure; $F_{e} = - eE_{V}
\sum_{p} p \partial f/\partial p = eE_{V}n_{q} = 2eE_{V}n_{s} +
eE_{V}n_{e}$ is the force of the electric field; $n_{e}$ is the density of
normal electrons; $n_{q} = n_{e} + 2n_{s}$ is the total density of
electrons; $F_{dis} = \sum_{p} p f_{1}/\tau $ is the dissipation force;
$F_{q} = \sum_{p} p \aleph$ is the quantum force.  $F_{q} = \sum_{p} p
df/dt + \sum_{p} p f_{1}/\tau $ during  $\Delta t_{qu}$

The quantum force $F_{qs}= \sum_{p} p df/dt$ acts directly on
superconducting pairs $\int_{\Delta t_{qu}} dt F_{qs} =  \sum_{p} p f_{qu}
- \sum_{p} p f_{cl} =  \Delta P_{s} =  (m/e)j_{p.c.}[1 + (Ls/l\lambda
_{0}^{2})n'_{s}] $ and $F_{qe} = \sum_{p} p f_{1}/\tau $ acts  on normal
electrons through the Faraday's voltage  $\int_{l}dl E = - (L/c) dI/dt$.
The dissipation force $F_{dis}$ strives to retain zero average velocity.
Therefore  $\Delta P_{e} = \int_{\Delta t_{qu}} dt F_{qe} =  \int_{\Delta
t_{qu}} dt F_{dis} = n_{e}(e/c) Lsj_{s}/l$.

Both $ P_{s}$ and $P_{n}$ return to initial values after the transition
$f_{qu} \rightarrow f_{cl}$ because of the dissipation force. After the
switching of a $l_{b}$ segment in the normal state with  $R_{bn} = \rho
_{n}l_{b}/s \neq 0$, when the resistance of other $l_{a}$ segment $R_{a} =
0$, a potential difference $V$ and a pressure difference is induced by the
deviation $\Delta n_{q}$ of the electron density from its equilibrium value
($\Delta n_{q} \ll n_{q}$). But $\int_{l}dl F_{p} = - \int_{l}dl \partial
(n_{q}<pv>)/\partial l = 0$ and $\int_{l}dl F_{e} = en_{q}\int_{l}dl E_{V}
= -en_{q}\int_{l}dl \nabla V = 0$

The order of $F_{p}$ and  $F_{e}$ magnitudes can be estimated by relations
$F_{p} \approx - <pv>\Delta n_{q}/\Delta l$ and $F_{e} \approx q^{2} n_{q}
\Delta n_{q}\Delta l = q^{2}/n_{q}^{-1/3} (\Delta l/n_{q}^{-1/3})^{2}
\Delta n_{q}/\Delta l$. Because $\Delta n_{q} \ll n_{q}$ the characteristic
length $\Delta l$ over which $n_{q}$ changes is much longer than the
distance between electrons: $\Delta l \gg n_{q}^{-1/3}$. In any metal $<pv>
\approx q^{2}/n_{q}^{-1/3}$ \cite{MetTheor}. Consequently, the force of the
pressure  $F_{p} \ll F_{e}$ is not important in our consideration.

The time of the $\Delta n_{q}$ appearance is very short because the
capacitance is very small. After this short time the $j_{sc}$ value is the
same in the superconducting $l_{a}$, $j_{sc} = j_{s} + j_{na}$, and in the
normal  $l_{b}$,  $j_{sc} = j_{nb}$, segments. The dissipation force acts
on superconducting pairs through the electric force  $\partial
P_{s}/\partial t = F_{e} = -2en_{s}\nabla V$ and $dj_{s}/dt =
(2e^{2}n_{s}/m)E_{a} =(2e^{2}n_{s}/m)(-\bigtriangledown V_{a} -
(Ls/c^{2}l)dj_{sc}/dt)$. The current of normal electrons $j_{na} = \rho
_{n}E_{a}$ in the $l_{a}$ segment and  $j_{nb} = \rho _{n}E_{b}$ in the
$l_{b}$ segment. Because $\int_{l}dl \nabla V = l_{a}<\nabla V_{a}> +
l_{b}<\nabla V_{b}> = < V_{a}> + <V_{b}> = 0$ the electric field $E_{a} =
-<V_{b}>/ l_{a} - (Ls/c^{2}l)dj_{sc}/dt)$ in the $ l_{a}$ segment and
$E_{b} = <V_{b}>/ l_{b} - (Ls/c^{2}l)dj_{sc}/dt)$ in the $ l_{b}$ segment.
At $l_{a} \gg  l_{b}$, when $j_{na} \ll j_{sc}$, $<V_{b}> \simeq
R_{bn}I_{sc} \simeq R_{bn}I_{s}\exp -t/\tau_{RL}$, where $\tau_{RL} =
(l_{a}/l + l_{a}\lambda _{0}^{2}/Lsn'_{s})L/R_{bn}$ is the decay time of
the current.

At $T \simeq  T_{cb} < T_{ca}$ only $l_{b}$ segment with lowest critical
temperature $T_{cb}$ is switched  in the normal state by the fluctuation.
In this case $\overline{R_{b}} \neq 0$,  $\overline{R_{a}} = 0$ and
$-\overline{<V_{a}>} = \overline{<V_{b}>} =L \overline{I_{s}}\omega
(l_{a}/l + l_{a}\lambda _{0}^{2}/Lsn'_{s}) $. Thus, not only the persistent
current $I_{p.c}$ but also the persistent voltage $V_{p.v.} =
\overline{<V_{b}>}$ can be induced by fluctuations in an inhomogeneous
loop. This result was published first in \cite{jltp98}. The possibility of
the persistent voltage is a direct consequence of the existence of the
non-chaotic Brownian motion at which  $\overline{F_{Lan}} =
\overline{F_{q}}  \neq 0$. The average force of the electric field
$\overline{F_{e}} = en_{q}\overline{E}$ should be not equal zero in an
inhomogeneous loop, in which the dissipation force $\overline{F_{dis}}$ has
different value in segments, because  $\overline{F_{q}}$ should be uniform
along the loop and according to (6)  $\overline{F_{e}}\simeq
\overline{F_{q}} - \overline{F_{dis}}$ (because $F_{p} \ll F_{e}$). In a
homogeneous loop $\overline{F_{e}} = \overline{F_{q}} - \overline{F_{dis}}
= 0$ because the switching probability of any segment is the same and
$\overline{F_{dis}}$ is uniform along the loop.

The inhomogeneous superconducting loop with $V_{p.v.} \neq 0$ is an
electric circuit in which the  $l_{a}$ segment with higher $T_{c}$ is a
power source, $W_{s} = \overline{<V_{a}>I_{sc}} \ < 0$, and the  $l_{b}$
segment with lower $T_{c}$ is a load, $W_{l} = \overline{<V_{b}>I_{sc}} \ >
0$. The power $W_{s}$ induced by the thermal fluctuation can not exceed
$(k_{B}T)^{2}/\hbar$ because the energy of fluctuation is $k_{B}T$ and the
frequency of switching $\omega < k_{B}T/\hbar$ in accordance with the
uncertainty relation. Consequently $V_{p.v.} = (R_{b}W_{l})^{0.5} <
k_{B}T_{c} (R_{b}/\hbar)^{0.5}$ in any case. $(k_{B}T)^{2}/\hbar \simeq
10^{-10} \ Wt$ at $T = 10 \ K$ and $(k_{B}T)^{2}/\hbar \simeq 10^{-8} \ Wt$
at $T = 100 \ K$. Therefore, at a real value $R_{b} = 1 \ \Omega$,  $
V_{p.v.} \ < 10^{-5} \ V = 10 \ \mu V$ for a low $T_{c}$ superconductor
with $T_{c} \approx 10 K$ and  $V_{p.v.} \ < 10^{-4} \ V = 100 \ \mu V$ for
a high $T_{c}$ superconductor with $T_{c} \approx 100 K$. These voltage
values are large enough to be measured experimentally.

The persistent voltage can be induced also in an inhomogeneous normal metal
mesoscopic loop \cite{NT32} in which the persistent current can exist
\cite{Kulik,IBM1991}. The mesoscopic loop, in which electrons are scattered
in only segment, is like the inhomogeneous superconducting loop considered
above. Superconducting condensate can be considered as a big particle which
is scattered on the normal loop segment like electrons are scattered on
impurities. In details the problem of the persistent voltage in an
inhomogeneous normal metal mesoscopic loop will be considered elsewhere.

The transition between $f_{cl}$ and $f_{qu}$ states can be induced not only
by the fluctuation but also by temperature change and by mechanical
interrupting and closing of the superconducting loop. In the first case the
loop can be considered as dc generator in which heat energy is transformed
in electric energy \cite{Jalta98}. In the second case the mechanical energy
is transformed to the electric energy.  In order to close the loop
interrupted by Josephson junction, an additional work $ \int db F_{q} =
\Delta b<F_{q}>$ should be expended because the energy is increased on
$E_{p} + E_{L} \approx E_{p} \approx (s/\lambda^{2})(\Phi_{0}^{2}/4\pi R)
(n - \Phi/\Phi_{0})^{2}$ at the $I_{sc}$ appearance.  The Josephson current
decreases exponentially with increasing of break width $b$ and has a
negligible value when $b$ exceeds some nanometers \cite{Barone}.
Consequently in order to close the loop at $n - \Phi /\Phi_{0} = 1/2$ the
quantum force, the average value of which equals $F_{q} \approx
(s/\lambda^{2})(\Phi_{0}^{2}/2l\Delta b)0.25$, should be overcome, where
$\Delta b \approx  10 \ nm$. At $l = 4 \mu m$, when $\Phi_{0}^{2}/2l
\approx 3 \ 10^{-20} \ J$,  $<F_{q}> \approx (s/\lambda ^{2}) \ 3 \
10^{-12} \ N$. This consideration shows that the wave function can have an
elasticity and that the quantum force can be connected with a real
classical force which can be measured.


\begin{thebibliography}{99}

\bibitem{little62} W.A.Little and R.D.Parks, {\it Phys.  Rev. Lett.} {\bf
9}, 9 (1962); {\it Phys. Rev.} {\bf 133}, A97 (1964).

\bibitem{repeat} H.Vloeberghs et al. {\it Phys. Rev.Lett.} {\bf 69},1268
(1992).

\bibitem{tink75} M.Tinkham, {\it Introduction to Superconductivity.} McGraw
-Hill Book Company (1975).

\bibitem{tinkham} M.Tinkham, {\it Phys. Rev.} {\bf 129}, 2413 (1963).

\bibitem{Isihara} A.Isihara, {\it Statistical Physics.} Academic Press, New
York - London, (1971)

\bibitem{Kulik} I.O.Kulik, {\it Pisma Zh.Eksp.Teor.Fiz.} {\bf 11}, 407
(1970) ({\it JETP Lett}. {\bf 11}, 275 (1970)); M.Buttiker, Y.Imry, and
R.Landauer, {\it Phys.Lett}. A {\bf 96}, 365 (1983).

\bibitem{IBM1991} E.M.Q.Jariwala et al. {\it Phys. Rev.Lett.} {\bf  86},
1594 (2001); V.Chandrasekhar et al. {\it Phys.Rev. Lett.} {\bf 67}, 3578
(1991); L.P.Levy et al. {\it Phys. Rev.Lett.} {\bf 64}, 2074 (1990)

\bibitem{Kohn57} W.Kohn and J.M.Luttinger, {\it Phys.Rev.} {\bf 108} 590
(1957).

\bibitem{MetTheor} A.A.Abrikosov, {\it Fundamentals of the Theory of
Metals}. North Holland, 1989.

\bibitem{jltp98} A.V. Nikulov and I.N. Zhilyaev, {\it J. Low Temp.Phys.}
{\bf 112}, 227 (1998)

\bibitem{NT32}  A.V. Nikulov, {\it Abstracts of 32 Russian Conference on
Low Temperature Physics}. Kazan, 3-6 October 2000, p. 100.

\bibitem{Jalta98}  A.V. Nikulov, {\it in Symmetry and Pairing in
Superconductors}, Eds.  M.Ausloos and S.Kruchinin, Kluwer Academic
 Publishers, Dordrecht, p.373 (1999).

\bibitem{Barone} A.Barone and G.Paterno, {\it Physics and Application of
 the Josephson Effect}. A Wiley - Interscience Publication, John Wiley and
 Sons, New York, 1982

\end{thebibliography}
\end{document}